\documentclass [10pt]{article}
\usepackage {graphicx}
\usepackage {color}

\title{Is Galaxy Distribution Non-extensive and Non-Gaussian? }
\author{Akika NAKAMICHI,$^1$ and
  Masahiro MORIKAWA,$^2$\\
  $^1$Gunma Astronomical Observatory, Takayama \\
Agatsuma, Gunma 377-0702, JAPAN\\
  $^2$Department of Physics, Ochanomizu University \\
2-1-1 Otsuka, Bunkyo, Tokyo 112-8610, JAPAN}

\def\LaTeX{L\kern-.36em\raise.3ex\hbox{a}\kern-.15em
    T\kern-.1667em\lower.7ex\hbox{E}\kern-.125emX}

\begin{document}

\label{firstpage}

\maketitle

%%%%%%%%%%%%%%%%%%%%
\begin{abstract}
%%%%%%%%%%%%%%%%%%%%
Self gravitating systems (SGS) in the Universe are generally thought to be non-extensive, and often show long-tails in various distribution functions. 
In principle, these non-Boltzmann properties are naturally expected from the peculiar property of gravity, long-range and unshielded.  
Therefore the ordinary Boltzmann statistical mechanics would not be applicable for these self gravitating systems (SGS) in its naive form.  
In order to step further, we quantitatively investigate the above two properties, non-extensivity and long-tails, 
by explicitly introducing various models of statistical mechanics.  
We use the data of CfA II South redshift survey and apply the count-in-cell method.  
We study four statistical mechanics, 
(1) Boltzmann, 
(2) Fractal, 
(3) R\'enyi, and 
(4) Tsallis, 
and use Akaike information criteria (AIC) for the fair comparison.
\end{abstract}

%%%%%%%%%%%%%%%%%%%%
\section{Introduction}
%%%%%%%%%%%%%%%%%%%%
The Long-range unshielded force gravity forms unique structures in the Universe called self-gravitating systems (SGS). 
Almost all the relevant structures such as stars, galaxies, clusters and super-clusters of galaxies, belong to this category. 

These SGS generally 
(a) show {\it non-extensive} properties, and 
(b) have {\it no absolute equilibrium state}; the systems never stop their evolution toward singularity.  

The former property (a) is apparent in principle.  
Actually, the mass density $\rho$ of iso-thermal SGS of size $r$ is given by 
\begin{eqnarray}
\rho=\frac{M}{4\pi r^3/3}=\frac{<v^2>}{4\pi G r^2/3}\propto r^{-2}
\label{rsv}
\end{eqnarray}
where we used the virial theorem: $K$=$-V/2$, 
where $K$ and $V$ are respectively the kinetic energy and the potential energy of the system.    
Since the general extensive structures have definite basic elements and therefore have constant density, 
the above scale-dependent density clearly shows the non-extensivity of SGS.  
This property reflects the long distance singularity of gravity.  
In other words, SGS cannot be divided into uniform independent ingredients (non-additive) in general.  

The latter property (b) is also apparent in principle.  
Actually, the partition function $Z$
of SGS in the temperature $T( = 1/\beta )$, again using the virial theorem, 
is given by 
\begin{eqnarray}
Z &=& \int_{}^{} {e_{}^{ - \frac{{\left( {K + V} \right)}}{{kT}}} dx_1^{} dp_1^{}  \cdots dx_N^{} dp_N^{} }   \\ \nonumber 
   &\propto& \int_{}^{} {e_{}^{ - \beta V/2} dx_1^{}  \cdots dx_N^{} }   \\ 
\nonumber
   &=& \int_{}^{} {\exp[{\beta G/2 \sum\limits_{i < j} {\left| {x_i^{}  - x_j^{} } \right|_{}^{ - 1} } }] dx_1^{}  \cdots dx_N^{} } 
\end{eqnarray}
which has an essential singularity at the short distance limit.  Therefore the absolute equilibrium state does not exist.  
This property reflects the short distance singularity of gravity.

Since the ordinary Boltzmann statistical mechanics is based on the extensivity and the existence of the absolute equilibrium state, the above properties of gravity make it inapplicable in principle.  

Despite the property (b), SGS frequently show {\it quasi}-equilibrium states which are often characterized by scaling properties; especially (c) the {\it long-tail} in various distribution functions are marvelous.  
This fact may suggest the existence of any underling fundamental statistical mechanics for SGS.  

There are many examples of the above properties (a) and (c) in SGS:  
The actual mean mass density of a structure is proportional to $r_{}^{ - 1.7} $
(de Vaucouleurs 1970)(Eq.(\ref{rsv}) is too simplified).  
Moreover in the extreme gravitating case of Black Hole, the `entropy' is not proportional to the mass $M$
but to the area $M_{}^2 $
.  
The property (c) is based on the absence of the characteristic scale in the Hamiltonian of SGS.  
The well known Holtzmark distribution of force $F$
acting on a single star in a uniform star cluster is exactly the stable Levy distribution with index 3/2 and behaves as $F_{}^{ - 5/2} $
in large force limit (Binney et al. 1987).  
The correlation function of galaxies and the clusters show power law behavior $\xi  \propto r_{}^{ - 1.86} $
(B\"{o}rner 2002).  

As explained in the above, the ordinary Boltzmann statistical mechanics cannot be applied to SGS in the naive form, and therefore any generalization or new statistical mechanics would be necessary in order to explain the quasi-equilibrium states of SGS.  

The necessary next step will be as follows, on which we will concentrate in this paper:  
\begin{enumerate}
\item[(A)]
To choose astronomical data which reflect the above properties (a), (b), and (c) in the fundamental level.  The data should be investigated not simply by the apparent form of the distribution functions but by internal structure of the theory that distinguish various statistical mechanics.  
\item[(B)]
To introduce a fair criterion which can specify the correct statistical mechanics among various proposed theories which have different number of free parameters.  
\end{enumerate}

For the purpose (A), we focus on the data of CfA II South (Huchra J., et al., 1999) galaxy distribution survey
\footnote{
We do not claim that this is the best available data, though it is uniform and large to some extent.  }.
Especially, we use count-in-cell method, in which the probability $f_V^{} \left( N \right)$
of finding a fixed number of galaxies in a fixed volume plays a central role. 	
For the purpose (B), we focus on the Akaike Information Criteria (AIC) (Akaike 1973)
\footnote{
We do not claim that this is the best available method; there are many extensions and generalizations of AIC.  These are unnecessarily complicated for the present purpose and therefore we choose the simplest version of AIC.  }.
This method will be most appropriate to compare different theories with different number of free fitting parameters.  

We have chosen the following four theories/models of statistical mechanics, classified according to the properties (a) non-extensivity and (c) long-tails.  (Table 1.)
 
\begin{itemize}
\item[]
(1) The Boltzmann statistical mechanics which is of course {\it extensive} from its construction.  The associated Boltzmann distribution function does not possess {\it long tail}.   
Since this theory alone never explain the CfAII data (see below), we introduce an extra parameter $b$ as in the reference by Saslaw W. C. \& Hamilton A. J. S. (1984).  
This parameter measures the possible deviation from the complete virial equilibrium. 
\item[]
(2)The (mono-)fractal-space model as a simple example of {\it non-extensive} 
theory with {\it short tail} distribution. 
\item[]
(3)The R\'enyi statistical mechanics as an appropriate example of {\it extensive} statistical mechanics with {\it long-tail} distribution.  
\item[]
(4)The Tsallis statistical mechanics (Nakamichi A., Joichi I., Iguchi O., Morikawa M., 2001), as a typical {\it non-extensive} theory.  
This has the same {\it long-tail} distribution as R\'enyi. 	
It should be noted that we use proper Tsallis statistical mechanics with Escote averaging, which should be distinguished from the older version of the normal averaging.  Both Tsallis and R\'enyi statistical mechanics posses an extra parameter $q$ which measures the deviation from the ordinary Boltzmann statistical mechanics; they reduce to the latter in the limit $q \to 1$
.  
\end{itemize}

Individual models/theories are explained in detail below.   

\begin{table}
\begin{tabular}{|p{40pt}|p{40pt}|p{40pt}|p{40pt}|}
\hline
& 
extensive& 
long-tail in distribution function& 
parameter (number) \\
\hline
Boltzmann& 
Yes& 
No& 
b (1)\\
\hline
Fractal (Boltzmann)& 
No& 
No& 
$\alpha$(1)
\\
\hline
R\'enyi& 
Yes& 
Yes& 
q, (s)  (1)\\
\hline
Tsallis& 
No& 
Yes& 
q, s (2) \\
\hline
\end{tabular}
\label{tab1}
\caption{
Four models/theories of statistical mechanics classified by (non)-extensivity of the model and the long/short tails of the associated distribution function.  
}
\end{table}

%%%%%%%%%%%%%%%%%%%%
\section{ Various models/theories of statistical mechanics}
%%%%%%%%%%%%%%%%%%%%

%%%%%%%%%%%%%%%%%%%%
\subsection[]{Boltzmann statistical mechanics}
%%%%%%%%%%%%%%%%%%%%

We consider the galaxy distribution which is supposed to obey the Boltzmann statistical mechanics with grand canonical ensemble.
In the ordinary Boltzmann statistical mechanics, the tail of the distribution function exponentially reduces. 
We phenomenologically generalize the theory and introduce the virial parameter $b$, which measures the deviation from the complete dynamical-equilibrium: 
\begin{equation}
p V = N T ( 1 - b ), 
\end{equation}
where $p$ denotes the pressure, $V$ the volume of the system, $N$ the number of galaxies contained in this volume, 
and $T$ the temperature defined by the velocity dispersion.

Usually, collision-less SGS attain the dynamical equilibrium before the thermal equilibrium: $\tau _{{\rm{thermal}}}^{}  \approx \frac{N}{{\ln N}}\tau _{{\rm{dynamical}}}^{}  \gg \tau _{{\rm{dynamical}}}^{} $ 
(see J. Binney et al. 1987 for detail).   
Therefore the parameter $b$ phenomenologically represents some class of quasi-equilibrium .  

The distribution function $f(0)$ is defined to be the probability of finding
no galaxy in any part of the volume $V$.  The explicit form is given by 
\begin{eqnarray}
f\left( 0 \right) = e_{}^{ - \bar N\left( {1 - b} \right)}  
= e_{}^{ - n\left( {1 - b} \right)4\pi r_{}^3 /3} 
\end{eqnarray}
where $n$ is the galaxy number density and ${\bar N}=nV$ (Saslaw 1984).  
This is the generating functional of the general function $f(N)$, which is defined to be the probability to find the $N$ galaxies in the fixed volume V.  
The general expression of $f(N)$ is given in the Appendix.  

From the observational data of CfA II South galaxy redshift survey, we calculate the probability $f(N)$ using the count-in-cell method.  
The best fit parameter $b$ is calculated to minimize the Akaike Information Criteria (AIC) for $f(0)$.  The comparison of the theory and observation is shown in Fig. 1.  
The diamonds with error bars are CfA II South Observations.  The void probability $f(0)$ calculated from the Boltzmann statistical mechanics is plotted by the broken line with the best fit parameter $b=0.523$.  
Quite a large deviation between the theory and the observation implies the inapplicability of the Boltzmann statistical mechanics even in the modified version.

\begin{figure}
\centerline{\includegraphics[width=84mm]{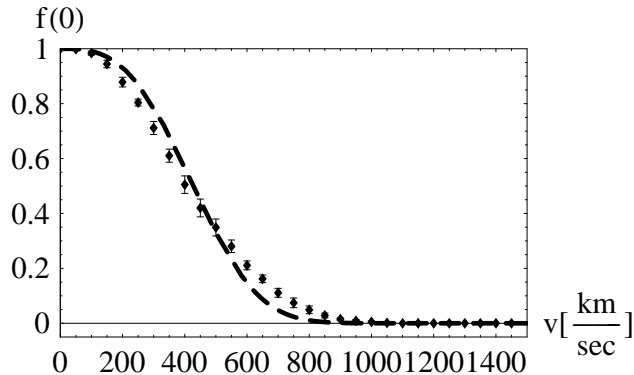}}
\label{fig1}
\caption{
The void probability $f(0)$ in the Boltzmann statistical mechanics.  
The horizontal axis is the distance scale in the unit of redshift [km/sec].  
The diamonds with error bars are CfA II South Observations.  Note that the number of the actual data points is about ten times larger than that shown in this figure for better visibility.  The void probability $f(0)$ calculated from the Boltzmann statistical mechanics is plotted by the broken line with the best fit parameter $b=0.523$.  
There exist quite a large deviation between the theory and the observation. }
\end{figure}

We fix the value of the parameter and with this value, general probabilities $f(N)$ are calculated and are compared with the CfA II South data in Fig. 2.  

\begin{figure}
\centerline{\includegraphics[width=84mm]{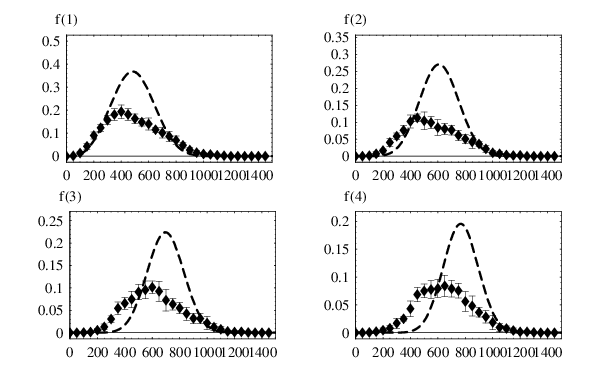}}
\label{fig2}
\caption{
The probability functions $f(1) - f(4)$ in the Boltzmann statistical mechanics. 
The horizontal axes are the distance scale in the unit of redshift [km/sec].  
The diamonds with error bars are CfA II South Observations. 
Theoretical predictions using Boltzmann statistical mechanics with the parameter $b =0.523$ are shown by the broken lines.
Note that the number of the actual data points is about ten times larger than that shown in this figure for better visibility.  
}
\end{figure}

%%%%%%%%%%%%%%%%%%%%
\subsection{Boltzmann statistical mechanics with fractal matter distribution}
%%%%%%%%%%%%%%%%%%%%

Various observational data suggest the idea that the matter distribution in the Universe is (multi-)fractal at least in some limited scale region (Kurokawa et al. 1999; 2000).  
In this section, we investigate a simple fractal model with the ordinary Boltzmann statistical mechanics.  In this model, the system is non-extensive in the sense that the simple addition of the two identical fractal distribution in the three dimensional space generally destroys the scaling property of the original fractal system.  
On the other hand, the tail of the distribution function exponentially reduces since the model is based on the ordinary Boltzmann statistical mechanics. 

We consider a mono-fractal matter distribution with the fractal dimension $\alpha$, and suppose the system obeys Boltzmann statistical mechanics with grand canonical ensemble, as in the previous section.

The void probability is simply given by $f\left( 0 \right) = e_{}^{ - \bar N}  = e_{}^{ - nV} $
, but the number $\bar N$
non-trivially depends on the scale $r$
in our fractal model.  
Thus $f(0)$ is given by 
\begin{eqnarray}
f\left( 0 \right) = e_{}^{ - \bar N}  = \exp \left[ { - n\frac{{\pi _{}^{\alpha /2} r_{}^\alpha  }}{{\Gamma \left( {1 + \alpha /2} \right)}}} \right]
\end{eqnarray}
This expression is shown in Fig.3 with the observational data of CfAII redshift survey.  By using the same method as in the previous section, we found the best fit value for the parameter $\alpha = 2.8821$. 

\begin{figure}
\centerline{\includegraphics[width=84mm]{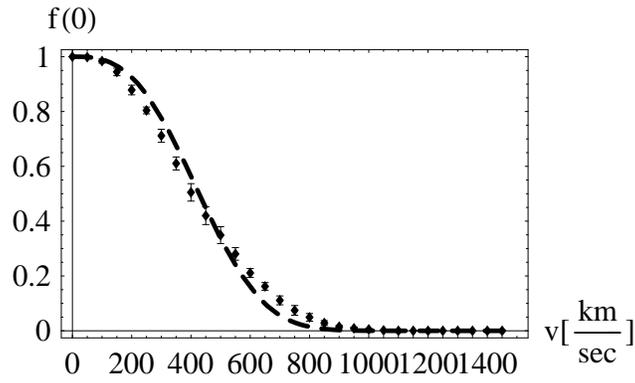}}
\label{fig3}
\caption{
The same as Figure 1. but in the fractal model based on the Boltzmann statistical mechanics.
The theoretical prediction with the best fit parameter $\alpha = 2.8821$ is shown by the broken line.  
There exist slight disagreement between the theoretical model and the observational data.}
\end{figure}

With this best fit parameter value, general probabilities $f(N)$ are calculated using the method in the Appendix, and they are compared with the CfA II South data in Fig. 4.  

\begin{figure}
\centerline{\includegraphics[width=84mm]{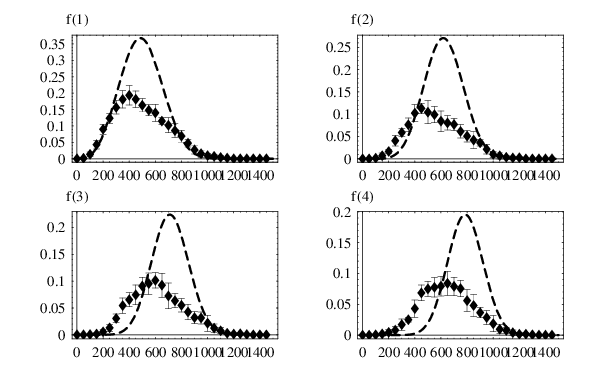}}
\label{fig4}
\caption{
Same as Figure 2. but in the fractal model with the Boltzmann statistical mechanics.  The parameter of fractal dimension is fixed as $\alpha  = 2.8821$.
}
\end{figure}

%However there also exist disagreement between the best fit theoretical model 
%and the observational data; accordingly, there is room for further investigation. 

%Hereafter, we will study alternative statistics instead of 
%above Boltzmann statistics.@

%%%%%%%%%%%%%%%%%%%%%%%%%%%%%%%%%%%%%%%%%%%%%%%%
\subsection{R\'enyi statistical mechanics} 
%%%%%%%%%%%%%%%%%%%%%%%%%%%%%%%%%%%%%%%%%%%%%%%%

In this section, we study R\'enyi statistical mechanics, which is extensive but the associated distribution function has {\it long-tail}.
R\'enyi statistical mechanics is a generalization of the ordinary Boltzmann theory by introducing a new form of entropy called R\'enyi entropy.  
This entropy is very similar to the information measure often used in multi-fractal models.  
However the use of this entropy as the starting point of the solid statistical mechanics is not clear.  
Therefore we concentrate on the application of this statistical mechanics in this paper without discussion of its physical foundation.  

The R\'enyi entropy is defined by 
\begin{equation}
S [p] = \frac{  \ln ( \sum_i {p_i}^q )}{1-q}, 
\end{equation}
where the parameter $q$ measures the deviation from the ordinary Boltzmann entropy.  Actually in the limit$q \to 1$
, this reduces to the Boltzmann entropy.  

When we compose two independent systems $A$ and $B$, the distribution function is given by 
$p_{i,j} [A, B]  =  {p_i}[A]  {p_j}[B] $,
and the composed entropy becomes 
\begin{equation}
 S_{A+B}  =  S_A  +  S_B, 
\end{equation} 
which clearly shows extensivity.   

Let us suppose that the galaxy distribution obeys the R\'enyi statistical mechanics with the grand canonical ensemble.  
Then the distribution function is given by 
\begin{equation}
p_{N, E}  =  [ 1 - \frac{1-q}{T} \{ E - \bar{E} - \mu (N - \bar{N}) 
\}     ]^{\frac{1}{1-q}}  
\end{equation}
which maximizes the above entropy. 

As is clearly seen from the above expression, the distribution function has a long-tail with power-law shape; a significant characteristic which distinguish this formalism and the Boltzmann formalism. 
	
Since the total entropy for $N$ galaxies, $S_N $, is simply the sum of $N$ entropies for a single galaxy:
$S_N = s N $, 
the Euler relation becomes the ordinary one:
\begin{equation}
S = \frac{ E + p V - \mu N }{T} .
\end{equation}
Using the above expression, the void probability becomes 
\begin{equation}
f ( 0 ) = \{ 1 + ( 1 - q ) N s \}^{-1} .
\end{equation}
Because the change of the parameter $s$ can be renormalized into the change of the parameter $q$, the probability is independent of $s$.
This fact naturally reflects the extensiveness of the present entropy.  

As in the previous section, using the count-in-cell method, 
we compare the theoretical model and the observational data of CfAII South.
\begin{figure}
\centerline{\includegraphics[width=84mm]{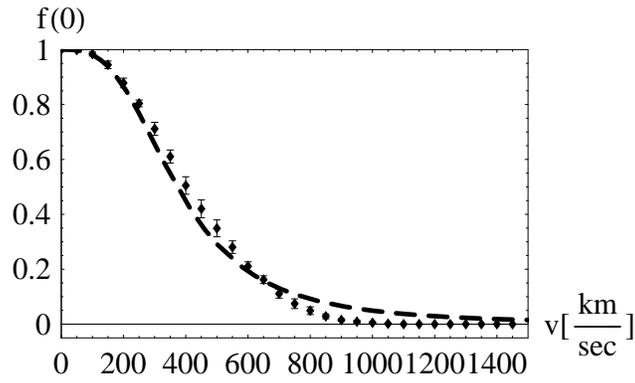}}
\label{fig5}
\caption{ 
Same as Figure 1. but in R\'enyi statistical mechanics.
Our best fit calculations with the parameter $q=-0.042$ and CfA II South Observations (diamonds with error bars) are plotted.
There exist slight disagreement between the theoretical model and the observational data.}
\end{figure}
The best fit value of the parameter $q$ is $q=-0.042 $. 
By fixing this value, we further obtain general probabilities $f(N)$ and they are compared with the CfA II South data.

\begin{figure}
\centerline{\includegraphics[width=84mm]{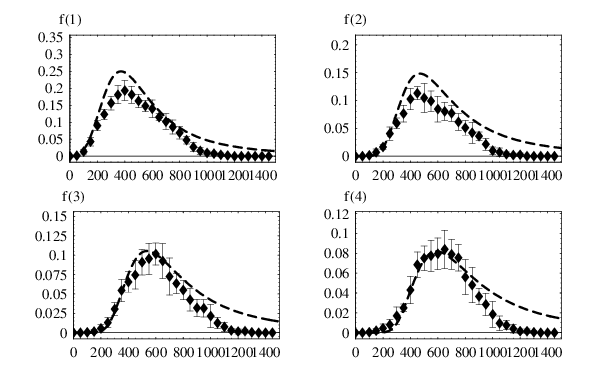}}
\label{fig6}
\caption{
Same as Figure 2. but in R\'enyi statistical mechanics using parameter 
$q =-0.042$.
}
\end{figure}

%%%%%%%%%%%%%%%
\subsection{Tsallis statistical mechanics} 
%%%%%%%%%%%%%%%

Tsallis C. (1988) proposed the non-extensive entropy 
\begin{equation} 
 S [P] = \frac{{(\sum_{i} {P_i}^{1/q})^{-q}} - 1}{1 - q}, 
\end{equation}
where the distribution function $P_i$ is the Escort distribution and is related with the bare distribution function $p_i$ by:
\begin{equation}
P_i = {p_i}^q / {\sum_{j=1}^{W} {p_j}^q} \equiv {p_i}^q / C .
\end{equation}
The Escort distribution is identified to be the physical distribution and is used to obtain observable averaging.  The distribution function is derived by maximizing the above entropy functional with appropriate constraints. 

 When we compose two independent systems $A$ and $B$, and for the distribution 
$p_{i,j} [A, B]  =  {p_i}[A]  {p_j}[B]  $, 
composed Tsallis entropy satisfies the non-extensive relation:
\begin{equation}
 S_{A+B}  =  S_A  +  S_B + ( 1 - q ) S_A S_B ,  
\end{equation}
as is easily calculated by using the above entropy form.  

Let us suppose the galaxy distribution obeys the Tsallis statistical mechanics with grand canonical ensemble. 
As in the case of fractal model and R\'enyi statistical mechanics, we do not need the virial parameter $b$ in Tsallis statistics. 

The distribution function 
\begin{equation}
p_{N, E}  = \frac{1}{\Xi_q}  [ 1 - \frac{1-q}{\tilde{T}} 
\{ E - \bar{E} - \mu (N - \bar{N}) \}     ]^{\frac{1}{1-q}}  
\end{equation}
maximizes Tsallis entropy. 
Here $\tilde{T} \equiv C T$ should be identified as the `temperature' of the system.  The factor $C$ appears everywhere in this theory.  

Note that the parameter $q$ measures the extent of deviation from the ordinary Boltzmann statistical mechanics; the distribution function reduces to the Boltzmann distribution for$q \to 1$.  

Tsallis distribution is identical to the R\'enyi distribution except the `temperature' and the normalization factor.  Therefore the Tsallis distribution also has the long-tail.  

Since the system is non-extensive, the total entropy of the $N$ galaxies should be calculated by using the composition law of the entropy given above.  The expression is given by  (Nakamichi A., et al., 2001)
\begin{equation}
S_N = \frac{ \{ 1 + ( 1 - q ) s \}^N - 1}{1-q} ,
\end{equation}
where the parameter $s$ is identified to be the entropy for a single galaxy. 

Then we obtain the generalized Euler relation \`{a} la Tsallis: 
\begin{equation} 
\frac{ \{ 1 + (1-q)S \}  \log [ 1 + (1-q)S ] }{1-q}= 
 \frac{E + p V - \mu N}{T} .
\end{equation}	

As in the previous two sections, using the count-in-cell method, probability of void becomes 
\begin{equation}
f(0) = \{ 1 + (1-q)s \}^{\frac{-N}{1-q}} [ 1 + 
   N \log \{ 1 + (1-q)s \}   ]^{\frac{q}{1-q}}  .
\end{equation}

\begin{figure}
\centerline{\includegraphics[width=84mm]{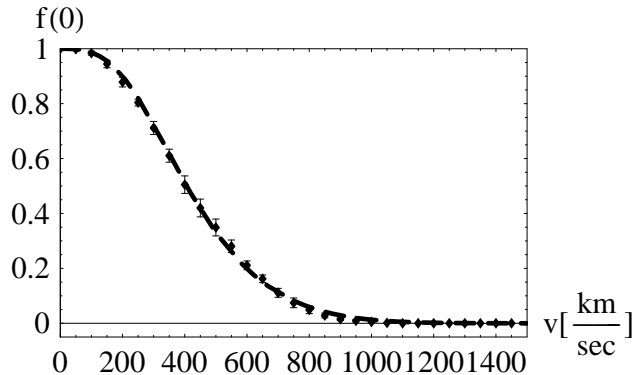}}
\label{fig7}
\caption{
Same as Figure 1. but in Tsallis statistical mechanics.
Our best fit ($q= -5.7 , s=0.16$) theory (broken line) and the CfA II South observations (diamonds with error bars) are plotted.
The fit seems to be the best among the four statistical mechanical models.  
}
\end{figure}

This prediction is compared with the observational data of CfAII South.
We found the best fit parameters $q=-5.7$ and $s =0.16$.  Almost complete fitting is remarkable.  

Further, we can calculate the general probability of finding $N$ galaxies in a given volume $V$, and compare with the CfA II South data in the figures 7,8.

\begin{figure}
\centerline{\includegraphics[width=84mm]{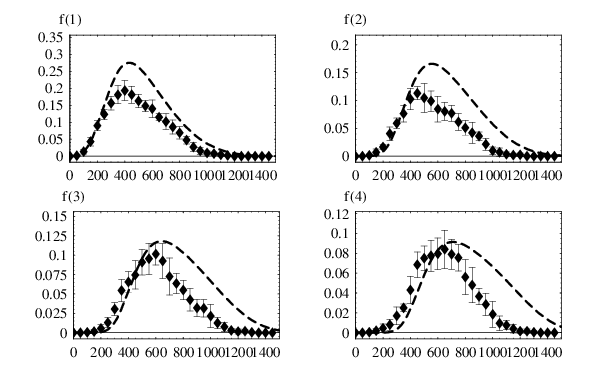}}
\label{fig8}
\caption{
Same as Figure 2. but in Tsallis statistical mechanics with Escote averaging, using parameters $q=-5.7$ and  $s =0.16$.
}
\end{figure}
We observe that for the higher order probability functions, the prediction by the Tsallis statistical mechanics is worse than the previous R\'enyi statistical mechanics, but is better than the Boltzmann statistical mechanics. 
%%%%%%%%%%%%%%%%%%%%%%%%%%%%%%%%%%%%%%

We have used the Escort averaging and not used the normal (old type) averaging in Tsallis statistical mechanics because the latter averaging is not a consistent theory.  
Tsallis originally proposed the non-extensive entropy using the normal averaging $p_i$ (see for example, Tsallis C., Mendes R. S., Plastino A. R., 1998)   
\begin{equation} 
 S [p] = \frac{(\sum_{i} {p_i}^q) - 1}{1 - q}, 
\end{equation}
where distribution function $p_i$ in normal averaging is determined so as to maximize the entropy functional. 

If we suppose the galaxy distribution obeys the above normal-averaging Tsallis statistical mechanics with grand canonical ensemble, the distribution function 
\begin{equation}
p_{N, E}  = \frac{1}{\Xi_q}  [ 1 - \frac{1-q}{\tilde{T}} 
\{ E - \bar{E} - \mu (N - \bar{N}) \}     ]^{\frac{1}{1-q}}  
\end{equation}
maximizes the Tsallis entropy. 
Here $\tilde{T} \equiv C T$ is the `temperature' of the system.
Also in this case, the distribution function approaches the ordinary Boltzmann distribution function for the parameter $q$ approaches to 1. 

The void probability is obtained in the same manner as the case of Escote averaging, but the expression has a slightly different form:  
\begin{equation}
f(0) = \{ 1 + (1-q)s \}^{\frac{-N}{1-q}} [ 1 + 
   N \log \{ 1 + (1-q)s \}   ]^{\frac{1}{1-q}}  .
\end{equation}
This time we have found the best fit values 
$q=-1000$ and $s =10^{200}$.  Even in this best fit case, there exists still significant difference between the theory and observational data.  The extremely large values above imply the absurdity of the theory.  
	
%%%%%%%%%%%%%%%%%%%%%%%%%%%%%%%
\section{Comparison of statistical mechanics using Akaike Information Criterion}
%%%%%%%%%%%%%%%%%%%%%%%%%%%%%%%

In order to calculate $f(N)$, we have studied several kinds of statistical mechanics, in which the number of free parameters are different.  
In general, a model with much more parameters can fit the observational data  better.  
However good physical theories must be simple and include the minimum number of free parameters. 
We need a fair measure to choose the correct statistical model to describe the galaxy distributions. 
One such measure will be the Akaike Information Criterion (AIC), (Akaike H.  1973; see for example, K. P. Burnham and D. R. Anderson 2002), which imposes a penalty on larger number of free parameters.  

AIC measure is given by 
\begin{eqnarray}
AIC &=& - 2 \times ({\rm maximum\ log\ liklihood}) \nonumber \\
	&+& 
	2 \times ({\rm number\ of\ parameters})
\end{eqnarray}
which is the unbiased estimator for expected mean log-liklihood.  
The first term on the right hand side tends to decrease as more parameters are introduced for better fitting, while the second term then becomes larger.
This is the tradeoff between the number of parameters and the variance.
Better model has smaller AIC measure.  

Here the ``parameters'' denotes the free parameters to be estimated by the maximum likelihood analysis. 
For the present argument, they are the variance estimated by the maximum likelihood analysis, and the other free parameters in each theoretical model.
	
Here the ``maximum log liklihood'' is calculated as follows. 
When we compare observational data sets $\left\{ {x_i ,\,\;y_i } \right\}_{1 \le i \le N} $
and the theoretical model $y\; = \;P(x)$, we assume that the errors (distance between the data and the theoretical model) distribute according to the Normal distribution:
$f\left( {x_i ,\,\;y_i } \right)\; = \;\frac{1}{{\sqrt {2\pi \sigma ^2 } }}\;e^{ - \frac{1}{{2\sigma ^2 }}\left( {y_i  - P(x_i )} \right)^2 }. $
Then the log-likelihood function becomes
\begin{eqnarray}
  \ell \left( \theta  \right)\; &\equiv& \;\sum\limits_{i = 1}^N {\ln \;f} \left( {x_i ,\,\;y_i } \right) \\ \nonumber 
   &=& \; - \frac{N}{2}\ln \left( {2\pi \sigma ^2 } \right) - \frac{1}{{2\sigma ^2 }}\sum {\left( {y_i  - P(x_i )} \right)^2 },  
\end{eqnarray}
where $\theta$ represents all the parameters in the model.  
The maximum likelihood estimate equations are 
\begin{eqnarray}
\frac{{\partial \ell \left( \theta  \right)}}{{\partial \theta }}\; &=& \;0, 
\\ \nonumber
\frac{\partial \ell \left( \theta  \right)}{\partial \sigma ^2 }\; &=& \; - \frac{N}{{2\sigma ^2 }} + \frac{1}{{2\sigma ^4 }}\sum\limits_{i = 1}^N {\left( {y_i  - P(x_i )} \right)^2 } \; = \;0.
\end{eqnarray}
The first equation determines the parameters $\hat \theta $.
The second equation determines the variance $\sigma ^2 _{} $
of the error to be estimated value  $\hat \sigma ^2 _{} $
as
\begin{eqnarray}
\hat \sigma ^2 \; = \;\frac{1}{N}\sum\limits_{i = 1}^N {\left( {y_i  - P(x_i )} \right)^2 }.  
\end{eqnarray}
Then the maximum log-likelihood function becomes 
$\ell \left( {\hat \theta } \right)\; = \; - \frac{N}{2}\; - \;\frac{N}{2}\;\ln 2\pi \hat \sigma ^2. $
Therefore AIC measure reduces to 
\begin{eqnarray}
AIC &=& ({\rm \# data\  points}) 
\times ( 1 + \log_{10} 2 \pi\ {\rm  (variance)} ) \nonumber \\
&+& 2 \times ( 1 + {\rm number\  of\  free\  parameters})
\end{eqnarray}

\begin{figure}
\centerline{\includegraphics[width=84mm]{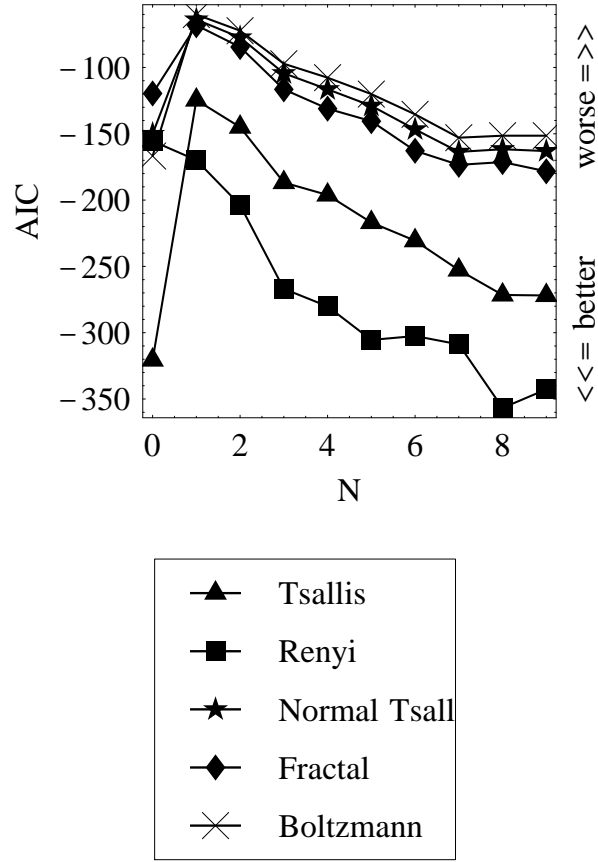}}
\label{fig9}
\caption{
AIC measure for probability of finding N galaxies in each theoretical model
for $f(N)$, ($0\leq  N\leq  9$).  
Vertical axis denotes the value of AIC. 
The smaller AIC, the theoretical model is better. 
Horizontal axis denotes the number of galaxies $N$ of $f\left( N \right)$.
AIC of Boltzmann statistical mechanics is plotted by 
crosses, 
fractal space model in Boltzmann statistical mechanics is plotted by 
diamonds, 
R\'enyi statistical mechanics is plotted by 
squares, 
Tsallis statistical mechanics with the Escote averaging is plotted by 
triangles, 
and the old type Tsallis statistical mechanics is plotted by 
stars.
}
\end{figure}

We have compared the AIC values for the probability of finding N galaxies in each theoretical model.
In the figure 9, such AIC values are plotted.  

For the void probability ($N=0$), we observe that the Tsallis statistical mechanics with the Escote averaging has a remarkably small AIC.
On the other hand for higher order probabilities (($1 \leq N$)), the R\'enyi statistical mechanics has much smaller AIC than the Tsallis with the Escote averaging.
Therefore only from these data, it is not possible to select the correct model among Tsallis and R\'enyi statistical mechanics. 
Therefore we cannot determin the (non)-extensivity of SGS.  
However, it is clear that the other models have large AIC values for all the probabilities.  
Therefore we may conclude that the long tail of the distribution function would be essential for describing the distribution of galaxies.

%%%%%%%%%
\section{Discussions} 
%%%%%%%%%  
We have studied two specific properties of SGS, 
non-extensivity of the entropy $S$, and 
the long-tails in the distribution functions, 
by explicitly introducing several models of statistical mechanics.  
By using the data of CfA II South redshift survey and the count-in-cell method, we have studied four statistical mechanics:
(1) Boltzmann, 
(2) Fractal, 
(3) R\'enyi, and 
(4) Tsallis.  
These statistical mechanics have been evaluated by the Akaike information criteria (AIC) for the fair comparison.

In our study, it has been essential that the expression of the probabilities 
$f(N)$ depends not only the distribution function but also the explicit form of the entropy.  
The long-tail property of distribution function affects the former dependence and the (non-)extensivity of entropy affects the latter dependence.  
Therefore we could distinguish the R\'enyi and Tsallis theories despite that they have almost the same form of distribution functions $P_{N, E}$ (see for example, Arimitsu T. \& Arimitsu N.  preprint 2002 ).

First we have seen that in Boltzmann statistical mechanics, it is unable to fit the CfAII South data of galaxy distribution even if we introduce the free parameter $b$ which measures the deviation from the complete dynamical-equilibrium, or the free parameter $\alpha$ which measures the fractal property of the space.   

Then we have investigated two statistical mechanics which have long tail distributions. 
Both the R\'enyi extensive statistical mechanics and the Tsallis non-extensive statistical mechanics are far better than the above two models based on Boltzmann statistical mechanics.  Therefore the long-tail in the distribution function will be essential for the correct statistical mechanics 
to describe SGS.  

On the other hand for the (non-) extensivity, we cannot have clear conclusions.  This is because the Tsallis non-extensive theory can fit $f(0)$ better than R\'enyi extensive theory, and the latter theory can fit $f(N)$ ($N \ge 1$
) better than the former theory.  
This incomplete conclusion may reflect that the size of the CfA survey data may still be small and much comprehensive survey would derive the complete answer to our question.  
We believe our method can in principle select the correct theory of statistical mechanics to describe SGS out of many candidate theories.  

\medskip
\noindent
{\bf Acknowledgement}
\noindent 
One of the authors (MM) would like to thank Prof. Hiroshi Hasegawa for very useful discussions and suggestions.

%%%%%%%%%%%%%%%%%%%%%%%%%%%%%%%%%%%%
\appendix
\section{Probabilities $P\left( N \right)$ and the generating functional}
%%%%%%%%%%%%%%%%%%%%%%%%%%%%%%%%%%%%

When we calculate the provability $P\left( N \right)$ of finding N galaxies 
in a given volume V, we use the expression 
\begin{equation}
\label{eq1}
P\left( N \right) \equiv \frac{( - n)^{N}}{N!}\frac{d^{N}}{dn^{N}}P(0)
\end{equation}
\noindent
where $n$ is the number density of galaxies. (White S. D. M. 1979)

	The proof of this expression is originally given by S. White in the above reference  
with cumbersome arguments. In this Appendix, we extract the essence of the 
proof and try to clarify the logic as much as possible. 

	For a general statistical variable $\phi \left( x \right)$, which is a field 
on three-dimensional space, the partition function is given by 
\begin{eqnarray}
\label{eq2}
 Z\left[ J \right] &=& \left\langle {\exp \left[ { - \int_{V}  {J\left( x 
\right)\phi \left( x \right)d ^{3} x} } \right]} \right\rangle \\ \nonumber
 &=& \exp \left\langle { - \int_{V}  {J\left( x \right)\phi \left( x 
\right)d ^{3} x} } \right\rangle _{c}  \\ \nonumber
 &\equiv& \exp W\left[ J \right]  
\end{eqnarray}
\noindent
where the brackets represent the functional integral of the field $\phi \left( x 
\right)$ (or the sum over possible functions $\phi \left( x \right))$, and 
$J\left( x \right)$ is a source field. 
The whole part of the cumulants or 
the connected correlation functions, denoted as the brackets with a suffix 
c, are \textit{defined} by the above equation. 

	We now specify the field $\phi \left( x \right)$ as (discontinuous, 
non-averaged) number density of galaxies: 
\begin{equation}
\label{eq3}
\phi \left( x \right) \to \sum\limits_{i = 1}^{\infty } {\delta  ^{3} 
\left( {x_{i}  - x} \right)} 
\end{equation}
\noindent
and accordingly the functional integration is reduced to the following form 
\begin{equation}
\label{eq4}
\left\langle \cdots \right\rangle \to \int_{V}  { \cdots \frac{d ^{3} 
x_{1}  }{V}\frac{d ^{3} x_{2}  }{V}\frac{d ^{3} x_{3}  }{V}} ...
\end{equation}
\noindent
since a distribution of galaxies at $\left\{ {x_{1}  ,x_{2}  ,x_{3}  
,...x_{n}  ,...} \right\}$ determines a field $\phi \left( x \right)$.
	 It is obvious that the probability of finding one galaxy within a small volume $d ^{3} x_{1}  $ around the space position $x_{1}  $ is $P\left( 
{x_{1}  } \right)d ^{3} x_{1}  = \left\langle {\phi \left( {x_{1}  } 
\right)} \right\rangle d ^{3} x_{1}  $, and the joint probability of 
finding one galaxy within a small volume $d ^{3} x_{1}  $ around the 
space position $x_{1}  $ \textit{and} the other within $d ^{3} x_{2}  $ around 
$x_{2}  $ is $P\left( {x_{1}  ,x_{2}  } \right)d ^{3} x_{1}  
d ^{3} x_{2}  = \left\langle {\phi \left( {x_{1}  } \right)\phi \left( 
{x_{2}  } \right)} \right\rangle d ^{3} x_{1}  d ^{3} x_{2}  $, 
and so on. 
	Much more useful quantities are the probability of finding no 
galaxy $P\left( 0 \right)$ within a fixed volume $V$. 
	Suppose the volume $V$ 
is divided into small pieces $\left\{ {v_{1}  ,...,v_{M}  } 
\right\}$, each of which is of order $v = V / M$. Then this void probability is 
\begin{eqnarray}
\label{eq5}
 P\left( 0 \right) &=& \left\langle {\prod\limits_{m = 1}^{M} {\left( {1 - 
\phi \left( {x_{m}  } \right)v_{m}  } \right)} } \right\rangle \\ \nonumber
 &=& \left\langle {\exp \left[ { - \sum\limits_{m = 1}^{M} {\phi \left( 
{x_{m}  } \right)v_{m}  } } \right] + O\left( {Mv ^{2} } \right)} 
\right\rangle  
\end{eqnarray}

Taking the limit $M \to \infty $ with fixed $V$, and therefore $O\left( 
{Mv ^{2} } \right) = O\left( {V ^{2} / M} \right)$, the void probability 
simply becomes 
\begin{eqnarray}
\label{eq6}
 P\left( 0 \right) &=& \left\langle {\prod\limits_{x \in V}  {\left( {1 - 
\phi \left( x \right)d ^{3} x} \right)} } \right\rangle  \\ \nonumber
 &=& \left\langle {\exp \left[ { - \int_{V}  {\phi \left( x \right)d ^{3} 
x} } \right]} \right\rangle  
 = Z\left[ 1 \right]  
\end{eqnarray}

	Similarly, the probability of finding one galaxy within a small volume 
$d ^{3} x_{1}  $ around the space position $x_{1}  $ \textit{and} finding no other 
galaxies is given by 
\begin{eqnarray}
\label{eq7}
 &&P\left( {x_{1\;}  ;\;0} \right)d ^{3} x_{1} \\ \nonumber
&=& \mathop {\lim }\limits_{M \to \infty } 
\left\langle {\phi \left( {x_{1}  } \right)v_{1}  \prod\limits_{m = 
2}^{M} {\left( {1 - \phi \left( {x_{m}  } \right)v_{m}  } \right)} } 
\right\rangle 
\\ \nonumber 
 &=& \mathop {\lim }\limits_{M \to \infty } \left\langle {\phi \left( 
{x_{1}  } \right)v_{1}  \exp \left[ { - \sum\limits_{m = 1}^{M} {\phi 
\left( {x_{m}  } \right)v_{m}  } } \right] 
+ O\left( {Mv ^{2} } 
\right)} \right\rangle 
\\ \nonumber
 &=& \left\langle {\phi \left( {x_{1}  } \right)\exp \left[ { - \int_{V}  
{\phi \left( x \right)d ^{3} x} } \right]} \right\rangle d ^{3} x_{1}  
\end{eqnarray}

General probability for N galaxies is given by 
\begin{eqnarray}
\label{eq8}
 &&P\left( {x_{1}  ,...,x_{N}  \;;\;0} \right)d ^{3} x_{1}  
...d ^{3} x_{N}  \\ \nonumber
&=& \left\langle {\begin{array}{l}
 \phi \left( {x_{1}  } \right)...\phi \left( {x_{N}  } \right)d ^{3} 
x_{1}  ...d ^{3} x_{N}  \times \\ 
 \quad \prod\limits_{\;\quad \;x \in V - \left\{ {x_{1} ,x_{2} , \cdots 
,x_{N} } \right\}\;}  {\left( {1 - \phi \left( x \right)d ^{3} x} 
\right)} \\ 
 \end{array}} \right\rangle \\ \nonumber
 &=& \left\langle {\phi \left( {x_{1}  } \right)...\phi \left( {x_{N}  } 
\right)\exp \left[ { - \int_{V}  {\phi \left( x \right)d ^{3} x} } 
\right]} \right\rangle d ^{3} x_{1}  ...d ^{3} x_{N}   
\end{eqnarray}

	On the other hand, the partition function can be expanded as 
\begin{eqnarray}
\label{eq9}
 Z\left[ J \right] &=& \sum\limits_{l = 0}^{\infty } {\frac{\left( { - 1} 
\right) ^{l} }{l!}\int   
{\begin{array}{l}
 \left\langle {\phi \left( {x_{1}  } \right)\phi \left( {x_{2}  } 
\right),...\phi \left( {x_{l}  } \right)} \right\rangle \\ 
 \quad \times J\left( {x_{1}  } \right)J\left( {x_{2}  } 
\right),...J\left( {x_{l}  } \right) \\
\quad \times dV_{1}  dV_{2}  ...dV_{l}  
 \end{array}} } \\ \nonumber
 &=& \sum\limits_{l = 0}^{\infty } {\frac{\left( { - 1} \right) ^{l} 
}{l!}\int   {\begin{array}{l}
 P\left( {x_{1}  x_{2}  ,...,x_{l}  } \right) \\ 
 \quad \times J\left( {x_{1}  } \right)J\left( {x_{2}  } 
\right),...J\left( {x_{l}  } \right) \\
\quad \times dV_{1}  dV_{2}  ...dV_{l}   
 \end{array}} }  
\end{eqnarray}

	In general, 
\begin{eqnarray}
\label{eq10}
 P\left( {x_{1}  ,...,x_{N}  } \right) &=& \left\langle {\phi \left( 
{x_{1}  } \right)...\phi    \left( {x_{N}  } \right)} \right\rangle 
\\ \nonumber
 &=& \left. {\left( { - 1} \right) ^{N} \frac{\partial  ^{N} Z\left[ J 
\right]}{\partial J\left( {x_{1}  } \right)...\partial J\left( {x_{N}  } 
\right)}} \right|_{J \to 0}   
\end{eqnarray}
\noindent
and
\begin{eqnarray}
\label{eq11}
 P\left( {x_{1}  ,...x_{N}  \;;\;0} \right) &=&  {\left\langle {\phi 
\left( {x_{1}  } \right)...\phi \left( {x_{N}  } \right)e ^{ - 
\int_{V}  \phi } } \right\rangle }  \\ \nonumber 
 &=& \left. {\left( { - 1} \right) ^{N} \frac{\partial  ^{N} Z\left[ J 
\right]}{\partial J\left( {x_{1}  } \right)...\partial J\left( {x_{N}  } 
\right)}} \right|_{J \to 1}   
\end{eqnarray}

	The probability of finding exactly $N$ galaxies with the volume $V$, 
using $P\left( {x_{1,}  x_{2}  ,...x_{N}  \;;\;0} \right)$ but not $P\left( 
{x_{1,}  x_{2}  ,...x_{N}  } \right)$, is given by 
\begin{eqnarray}
\label{eq12}
 P\left( N \right) &=& \frac{1}{N!}\int_{V_{1}  }  {...\int_{V_{N}  }  
{P\left( {x_{1,}  x_{2}  ,...x_{N}  \;;\;0} \right)} } \\ \nonumber
 &=& \left. {\frac{\left( { - 1} \right) ^{N} }{N!}\int_{V_{1}  }  
{...\int_{V_{N}  }  {\frac{\partial  ^{N} Z\left[ J \right]}{\partial 
J\left( {x_{1}  } \right)...\partial J\left( {x_{N}  } \right)}} } } 
\right|_{J \to 1}  
\end{eqnarray}

If we factor out the mean number density $n$ from the density field $\phi 
\left( x \right)$ as $\phi \left( x \right) = n\left( {1 + \delta \left( x 
\right)} \right)$, where $\delta \left( x \right)$ denotes deviation from the 
average, then we have 
\begin{equation}
\label{eq13}
\int_{V}  {\left. {\frac{\partial Z\left[ J \right]}{\partial J\left( x 
\right)}} \right|_{J \to 1}  } = n\frac{\partial Z\left[ 1 
\right]}{\partial n}
\end{equation}
\noindent
and similarly 
\begin{equation}
\label{eq14}
\left. {\int_{V_{1}  }  {...\int_{V_{N}  }  {\frac{\partial 
^{N}Z\left[ J \right]}{\partial J\left( {x_{1} } \right) \cdots \partial 
J\left( {x_{N} } \right)}} } } \right|_{J \to 1}  = n^{N}\frac{\partial 
^{N}Z\left[ 1 \right]}{\partial n^{N}}
\end{equation}

	Using this form and Eq.(\ref{eq6}) for Eq.(\ref{eq12}), we finally have 
\begin{equation}
\label{eq15}
P\left( N \right) = \frac{\left( { - n} \right) ^{N} }{N!}\frac{\partial 
 ^{N} }{\partial n ^{N} }P\left( 0 \right)
\end{equation}
\noindent
as desired.

\end{document}